\documentclass{article}

\usepackage{calc}
\usepackage{amssymb}
\usepackage{amstext}
\usepackage{amsmath}
\usepackage{amsthm}
\usepackage{multicol}
\usepackage{array}
\usepackage{url}
\usepackage{xcolor}
\usepackage{soul}
\usepackage{graphicx}

\usepackage{algorithm}
\usepackage{algorithmic}
\usepackage[accepted]{icml2014}

\icmltitlerunning{Person Identification Based on Hand Tremor Characteristics}

\begin{document}

\twocolumn[ \icmltitle{Person Identification Based on Hand Tremor Characteristics}
\icmlauthor{Oana Miu}{omiu@fotonation.com} \icmladdress{Fotonation Romania\\
                Address Calea Serban Voda, 133, Bucharest, Romania }

\icmlauthor{Adrian Zamfir}{azamfir@fotonation.com} \icmladdress{Fotonation Romania\\
                Address Calea Serban Voda, 133, Bucharest, Romania }

\icmlauthor{Corneliu Florea}{cflorea@fotonation.com} \icmladdress{Fotonation Romania\\
                Address Calea Serban Voda, 133, Bucharest, Romania }
                \icmladdress{Image Processing and Analysis Laboratory\\
                University ``Politehnica" of Bucharest, Romania,
                Address Splaiul Independen\c{t}ei 313}

\icmlkeywords{tremor recognition, inertial sensors}

\vskip 0.3in

]

\begin{abstract}
A plethora of biometric measures have been proposed in the past. In this paper we introduce a new
potential biometric measure: the human tremor. We present a new method for identifying the user
of a handheld device using characteristics of the hand tremor measured with a smartphone built-in
inertial sensors (accelerometers and gyroscopes). The main challenge of the proposed method is
related to the fact that human normal tremor is very subtle while we aim to address real-life scenarios.
To properly address the issue, we have relied on weighted Fourier linear combiner for retrieving
only the tremor data from the hand movement and random forest for actual recognition. We have
evaluated our method on a database with 10 000 samples from 17 persons reaching an accuracy of 76\%.
\end{abstract}

\section{Introduction}
\label{sec:introduction}

Biometric recognition refers to an automatic identification of individuals using metrics derived from
their physiological and/or behavioral characteristics \cite{jain2000}. The purpose of such a system
is to ensure that sensitive applications, such as computer systems security, mobile phones, credit
cards, secure access to buildings - are accessed only by a legitimate user and by nobody else. By
using biometric recognition, a system can identify a person based on ``who she/he is" rather than
``what she/he has" (card, token, key) or ``what she/he knows" (password, PIN). While biometric
systems may get extremely accurate even for large scale identification applications (as for
instance the US-VISIT program based on fingerprint recognition) usually the performance comes with the
cost of intrusive evaluation which leads to a low acceptability from people.

In the recent years we note the expanding use of smartphones in our daily life. In 2013,
smartphones sales have surpassed in U.S. the basic mobile phones  \cite{ipsos2013} and in 2015 have 
become the most employed Internet access device \cite{bosomworth2015}. Smartphones, if left unsecured, can
grant access to the user's private information: email correspondence, address book, any unsecured
documents and potentially banking accounts. Lock screens have been proven inefficient for a user
that is concerned with security and some smartphone models include fingerprint readers.

In this paper we would like to point the attention of the reader to the human involuntary hand tremor
as a new biometric measure. Furthermore, as currently most smartphones are delivered with inertial
sensors capable of measuring it, its applicability is immediate and we will show that the reachable
discrimination power is at least sufficient for low-security scenarios.

Tremor is a behavioral biometric falling in the same category as gait \cite{little1998},
\cite{nixon2006}:  while for limited amounts of time it is reliable, it may not remain invariant
especially over a long period of time, due to major injuries involving hand joints, certain
medications, inebriety, muscle tiredness or weakness, normal aging, stress, anxiety or fatigue
\cite{andrade2013}. As recent studies \cite{veluvolu2011} have shown that tremor can be efficiently
distinguished from voluntary movement, our idea is to use tremor based features to identify the person 
that holds the smartphone.

We propose a new biometric system based on involuntary tremor detection using inertial sensors
(accelerometer and gyroscope) that are already integrated into smartphones. This system can be used
for low security applications like automatically unlocking the smartphone only if it is held by the
recognized owner. While the here presented study shows high discrimination power over a limited
number of persons, we cannot be sure that tremor fingerprint is unique amongst millions of persons.

The remainder of the paper is constructed as follows: in Section \ref{Sect:RelatedWork} we discuss
relevant prior work, in section \ref{Sect:Methodology} we present the proposed methodology, while
in section \ref{Sect:Implement} to discuss the achieved results. Section \ref{Sect:Discussion} is
dedicated to conclusions and discussions.


\section{Related Work}
\label{Sect:RelatedWork}

\paragraph{Hand Tremor Studies.} The human tremor is defined as ``a rhythmic and involuntary oscillation of
a body part, caused by reciprocal innervations of a muscle, which leads to repetitive contractions"
\cite{mansur2007}. The human tremor can be categorized in two main classes: resting tremor (which
can be noticed when the muscles are not contracted) and action tremor \cite{andrade2013}. The
action tremor manifests during a voluntary muscle contraction. The action tremor encompasses
postural, kinetic, intentional, task-specific and isometric tremor and is the type of tremor that
is the most likely to appear when using a smartphone. Another classification of the tremor types
takes into account its nature \cite{mansur2007}: physiological tremor (which is present in all
healthy people) and pathological tremor (associated with various diseases or conditions such as
Parkinson disease). Multiple studies summarized by Mansur et al. \cite{mansur2007} concluded that
physiological tremor has most of the energy in the [7-12] Hz domain while pathological one has many
components in lower ranges. In the current study we have only included persons known not to be
suffering from any tremor-related conditions.

As all humans exhibit a form of tremor, its study has received sufficient attention.  Tremor
compensation is important in microsurgical applications \cite{veluvolu2011} where involuntary
movements needed to be counteracted. The majority of the existing techniques rely on low-pass
filtering approaches which are successful in compensating tremor, but the inherent time delay is a
major drawback. To overcome this problem, adaptive filtering approaches have been developed and are
well-suited for tremor estimation as they can adapt to the changes both in frequency and in
amplitude of the tremor signal. Fourier Linear Combiner (FLC) \cite{vaz1994} is an adaptive filter
that operates by adaptively estimating the Fourier coefficients of the known frequency model
according to the least mean square (LMS) optimization algorithm. Weighted frequency Fourier Linear
Combiner (WFLC) is an adaptive algorithm which models any quasi-periodic signal as a modulating
sinusoid and tracks its frequency, amplitude and phase which means it incorporates the frequency
adaptation procedure into FLC and can be successfully used for adaptive tremor cancellation
\cite{riviere1998}.

Another aspect of major interest was the automatic recognition of one of the tremor categories. The
most popular division is between physiological tremor, essential tremor and parkinsonian tremor.
High separation rates are reported by \cite{jakubowski2002}, which relies on a multi-layer
perceptron classification of features derived from high order statistics, or by \cite{soran2012}
which feeded the filter output in a Support Vector Machine (SVM).

In parallel, due to their inclusion in smartphones, a multitude of applications are based  on
inertial sensors. For instance, let us refer to the work of Sindelar and Sroubek
\yrcite{sindelar2013} which aimed to remove the camera shake without hardware stabilization or to
\cite{siirtola2012} which has aimed at user activity recognition.

\paragraph{Biometric Measures.} Currently a multitude of biometric measures were proposed and evaluated
in previous academic works. In the seminal review on the topic, Jain et al. \yrcite{jain2004}
discussed the following measures: DNA, ear (shape and structure of the cartilaginous tissue), face,
facial, hand and hand vein infrared thermograms, fingerprint, gait, hand and finger geometry,
iris, keystroke, odor, palmprint, retinal scan, signature and voice. In a more recent review of
biometric measures, Unar et al. \yrcite{unar2014} have added finger knuckle print, tongue print and also
point to the so called soft biometrics (i.e ``characteristics that lack distinctiveness and
permanence because they are most common among humans").

The use of a hand tremor as a biometric measure, to our best knowledge was not discussed previously
in the academic literature, but only named in the intellectual property domain \cite{liberty2007}.
Regarding this work we must point out that, due to the specificity of the publication domain, it
does not report any actual results and only discuss general approachable designs, out of which in
the preferred implementation also includes a laser pointer for identifying the user that holds
a remote control or a mouse.

Thus we argue that the work proposed here, as it details not only the idea but also the system
used for implementation and reports results that are obtained in real life scenarios, opens the
path for a new direction of research with great potential.


\section{Methodology}
\label{Sect:Methodology}

\begin{figure*}[t]
    \begin{center}
    \includegraphics [width=0.70 \textwidth]{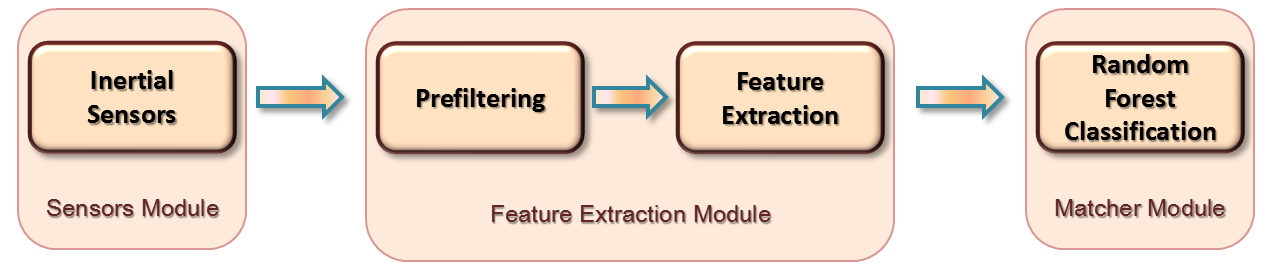}
    \caption{The stages of the proposed hand tremor analysis system.}
    \label{fig:method}
    \end{center}
\end{figure*}

The proposed hand tremor analysis system (shown in Figure \ref{fig:method}) contains the first
three typical \cite{jain2004}, \cite{unar2014} modules of a biometric system.

The first module regards data collecting from the inertial sensors which are built-in the
mobile phone.

The second stage includes two steps: complex movement filtering to extract the tremor data and the actual 
feature extractor that is used to obtain the attributes employed for tremor classification.

Finally, in the matcher module, which in fact is a decision making module, the identity of the
presumed user is established. The actual implementation relies on a Random Forest classifier.

The sensors' data acquisition and the prefiltering method are discussed in sections
\ref{Sect:Methodology}.1 and \ref{Sect:Methodology}.2. The feature extraction process is explained
in section 3.3 and the classification is discussed in section 3.4.


\subsection{Sensors}
Our tests were focused on smartphones available on the market and on the inertial sensors delivered
implicitly: 3-axes gyroscopes and accelerometers. Each of the three spatial axis (X,Y,Z) movement is
recorded by an accelerometer (which measures translational acceleration) and by a gyroscope (which
measure rotational velocity).

For actual recording, we have developed an application that records the gyroscope and accelerometer
data at a sample rate of 100 Hz. While this was the best choice provided by the API we have used we point 
that any higher frequency, while it comes with cost of more resources, may be beneficial for system accuracy.

To eliminate unwanted movements from the beginning and end of each recording session we have removed the
first and last 100 milliseconds, which may be affected by the phone state transition. We divide each
signal axis recording into one second windows. The data from one window will form a sample from our
data set.


\subsection{Prefiltering}

To remove the signal noise we rely on the WFLC algorithm for filtering. As previously mentioned,
the WFLC is an adaptive filter that can adapt to a quasi-periodic signal of an unknown frequency,
amplitude and phase. The WFLC algorithm is an extension of the standard FLC algorithm that showed
to lead to superior performance when faced with a signal that displays an oscillatory pattern but
with a time-varying period as WFLC may estimate the signal value at non-fixed frequencies
\cite{riviere1998}. In the next paragraphs we will recall the WFLC algorithm main steps.

The input of the WFLC is the inertial sensor signal $(s_{k})$  as measured at the sampling moment
$k$.  The adaptive vector $w_{k}$ estimates the signal amplitude and phase and $\omega_{0_{k} }$,
the frequency of the signal. Depending on the instantaneous difference, $\epsilon_{k}$,
WFLC adjusts, at each iteration, $\omega_{0_{k} }$ and $w_{k}$. The problem is a typical convex
optimization one and it is solved by the adjustment of the standard least mean square (LMS)
algorithm.

The state vector $x_k = [ x_{1_k } . . . x_{2M_k}  ]^T$ used by the WFLC algorithm is composed of
the sine and cosine functions computed using the frequency weight $w_{0_k}$ and $M$ is the order of
the Fourier series representing the measured signal $s$. $\mu_{0}$, $\mu$,  $\mu_{b}$ are the
frequency, amplitude and bias adaptation gains, respectively.

\begin{equation}
x_{rk}=\left\{ \begin{array}{rl}
\sin\left({r\sum\limits_{t=0}^{k}}\omega_{0_{t} }\right)   & \mbox{  $1\leq r \leq M$} \\
\cos\left({(r-M)\sum\limits_{t=0}^{k}}  \omega_{0_{t} }\right)   & \mbox{ $M+1 \leq r \leq 2M$}
\end{array} \right.
\end{equation}

The update is:

\begin{equation}\label{eq2}
\epsilon_{k}= s_{k}-w_{k}^T x_{k}+\omega_{b_{k} }
\end{equation}

  $\omega_{b_{k}} $ is introduced in the computation of $\epsilon_{k}$ to estimate and
remove  the bias present in the signal \cite{veluvolu2011}, due to possible low frequency
components and/or drift.

The initialization is:

\begin{equation}\
 \omega_{0_{k+1}} =\omega_{0_{k}} +2\mu_0 \epsilon_k \sum_{r=1}^M r(w_{r } x_{M+r }-w_{M+r} x_{r })
\end{equation}
  while the unknown frequency and adaptive weight updates are:

\begin{equation}
w_{k+1}=w_k+2\mu x_k \epsilon_k
\end{equation}

\begin{equation}
\omega_{b_{k+1} }=\omega_{b_k }+2\mu_b \epsilon_k
\end{equation}

The order of the Fourier series $M$ used for representing the measured signal $s$, was set to
$M=5$.  We have found that $\mu_0=10^{-5}$, $\mu=0.3,$ $\mu_b=2.5$ \textperiodcentered $ 10^{-8} $
and $w_0=2$ produced the best estimate for the voluntary movement.

As the accelerometers and gyroscopes have three axes $X$, $Y$ and $Z$, we filter using WFLC each of the
axes of accelerometer and gyroscope data.

We then compute the acceleration magnitude as:

\begin{equation}
A_{norm}=\sqrt{A_x^2+A_y^2+A_z^2}
\end{equation}

The gyroscope magnitude is computed similarly and is used for the tremor features extraction.


\subsection{Feature extraction}

Following the previous works on classifying the tremor into physiological categories we note that
the main characteristics are related to favored frequencies \cite{mansur2007}, \cite{andrade2013}.
Thus, the extracted features are focused on spectral description of the acquired signals.

For the actual implementation we relied on LibXtract \cite{bullock2007} and extracted 12 features
in both time and frequency domains (listed Tables \ref{tab:features1} and \ref{tab:features2}) for
each of the gyroscope and accelerometer axis and magnitude.

Also, as tremor is a quasi-periodic movement, we have considered that the spectral density
estimation can produce more information on the rhythmic movement. Spectral density characterizes the
frequency content of a signal and detects any periodicities in the data, by observing peaks at the
frequencies corresponding to these periodicities. The simplest technique to estimate the spectrum
is the periodogram, given by the modulus squared of the discrete Fourier transform
\cite{stoica2005}. We have extracted for the periodogram (i.e. discrete approximation of the power
spectral density) some of the features that we have used for the frequency domain and added the top
three frequencies, which define the tremor as showed in Table \ref{tab:features3}.

Overall it resulted in (3 axis + magnitude) $\times$ (2 sensors) $\times$ (22 features) = 176
features for each sample.

\begin{table}[t]
  \centering
    \caption{List of Time Domain Features. Vector $\mathbf{x}$ is the time domain representation of data.
       $N$ is the number of elements in $\mathbf{x}$.}\label{tab:features1}
    \begin{tabular}{|m{1.5cm}|m{5.5cm}|}
        \hline
        \bf Feature Name   & \bf Description                                                     \\ \hline
        Mean               & $\bar{x} = \frac{1}{N} \sum\limits_{i=1}^{N} x(i) $                 \\ \hline
        Standard Deviation & $\sigma = \sqrt{\frac{1}{N-1} \sum\limits_{i=1}^{N}(x(i)-\bar{x})^2}$ \\ \hline
        Average deviation  & $D_x = \frac{1}{N} \sum\limits_{i=1}^{N} |x(i)-\bar{x}|$            \\ \hline
        RMS Amplitude      & $A = \sqrt{\frac{1}{N} \sum\limits_{i=1}^{N}(x(i))^2} $             \\ \hline
        Highest Value      & $H=(\max{x(i)}\vert_{i=1\dots N} )$                               \\ \hline
    \end{tabular}
\end{table}

\begin{table}[h]
\centering
    \caption{List of Frequency Domain Features. Vector $\mathbf{y}$ is the frequency representation of data.
     Vectors $\mathbf{y_m}$ and $\mathbf{y_f}$ hold the magnitude coefficients and bin frequencies respectively.
     $N$ is the number of elements in $\mathbf{y_m}$ and $\mathbf{y_f}$.}\label{tab:features2}
    \begin{tabular}{|m{1.5cm}|m{5.5cm}|}
        \hline
        \bf Feature Name            & \bf Description                                                                                                         \\ \hline
        Spectral Standard Deviation & $\sigma_s = \sqrt{\left( \sum\limits_{i=1}^{N} \left(y_f (i)\right)^2 y_m (i)\right)/\left(\sum\limits_{i=1}^{N} y_m(i)\right) } $ \\ \hline
        Spectral centroid           & $C_s = \left( \sum\limits_{i=1}^{N} y_f (i) y_m (i)\right)/\left(\sum\limits_{i=1}^{N} y_m(i)\right)  $               \\ \hline
        Spectral skewness           & $\gamma_s = \left( \sum\limits_{i=1}^{N} (y_m (i)-C_s)^3\right)/\sigma_s^3$                                                        \\ \hline
        Spectral Kurtosis           & $\beta_s = \left( \sum\limits_{i=1}^{N} (y_m (i)-C_s)^4\right)/\sigma_s^4 - 3$                                                     \\ \hline
        Spectral Crest                 & $CR_s=\left(\max{y_m(i)}\vert_{i=1\: to\: N} \right)/C_s$                                                                          \\ \hline
        Irregularity-K              & $IK_s=\sum\limits_{i=1}^{N-1} {\left|y_m(i)-\frac{y_m(i-1)+y_m(i)+y_m(i+1)}{3}\right|}$                                            \\ \hline
        Irregularity-J              & $IJ_s=\frac{\sum\limits_{i=1}^{N-1} \left(y_m (i)-y_m(i+1)\right)^2}{\sum\limits_{i=1}^{N-1} \left(y_m (i)\right)^2}$                         \\ \hline
    \end{tabular}
\end{table}

\begin{table}[t]
    \caption{List of Power Spectral Density Features. Vector $\mathbf{p}$ represents the periodogram
    of the data. Vectors $\mathbf{p_m}$ and $\mathbf{p_f}$ hold the magnitude coefficients and bin frequencies,
    $N$ is the number of elements in $\mathbf{p_m}$ and $\mathbf{p_f}$.}\label{tab:features3} \centering
    \begin{tabular}{|m{1.7cm}|m{5.4cm}|}
        \hline
        \bf Feature Name                  & \bf Description                                                                                                                                         \\ \hline
        Periodogram Standard Deviation    & $\sigma_p = \sqrt{\left( \sum\limits_{i=1}^{N} (p_f (i))^2 p_m (i)\right)/\left(\sum\limits_{i=1}^{N} p_m(i)\right) } $                                                                     \\ \hline
        Periodogram centroid              & $C_p = \left( \sum\limits_{i=1}^{N} (p_f (i)) p_m (i)\right)/\left(\sum\limits_{i=1}^{N} p_m(i)\right)  $                                                                                   \\ \hline
        Periodogram skewness              & $\gamma_p = \left( \sum\limits_{i=1}^{N} (p_m (i)-C_p)^3\right)/\sigma_p^3$                                                                                               \\ \hline
        Periodogram Kurtosis              & $\beta_p = \left( \sum\limits_{i=1}^{N} (p_m (i)-C_s)^4\right)/\sigma_p^4 - 3$                                                                                            \\ \hline
        Periodogram Crest                     & $CR_p=\left(\max{p_m(i)}\vert_{i=1\: to\: N} \right)/C_s$                                                                                                          \\ \hline
        Periodogram Irregularity-K        & $IK_p=\sum\limits _{i=1}^{N-1} {\left|p_m(i)-\frac{p_m(i-1)+p_m(i)+p_m(i+1)}{3}\right|}$                                                                \\ \hline
        Periodogram Irregularity-J        & $IJ_p=\frac{\sum\limits_{i=1}^{N-1} (p_m (i)-p_m(i+1))^2}{\sum_{i=1}^{N-1} (p_m (i))^2}$                                                                       \\ \hline
        Top three periodogram frequencies & $[H_1 ;  H_2 ;  H_3]=[(\max{p_m(i)}\vert_{i=1\dots N_1} )\newline (\max{p_m(i)}\vert_{i=N_1\dots N_2} ) \newline (\max{p_m(i)}\vert_{i=N_2 \dots N} )]$ \\ \hline
    \end{tabular}
\end{table}


\subsection{Matcher Module}
In the current evaluation we have focused our attention on an identification scenario: the system aims
to recognize the individual by searching the templates of all the users in the database for a
match. From a classification point of view, the system should be trained and tested over
categorical variables; thus a  machine learning step is deemed.

While in the later period the deep networks \cite{lecun2015} seem to provide the best accuracy, yet
they need large training sets and even with latest progress \cite{ba2014} they still require
significant resources while testing. From the rest of classifiers, following the large scale study
by \cite{fernandez2014} indicates the family of aggregated decision trees as the best performing.
This family also has the advantage of having a good generalization error while testing and being robust
to various stresses.

As we will further show in the next section the best result was achieved with a Random Forest Classifier
\cite{breiman2001}.


\section{Implementation and Results}
\label{Sect:Implement}

\subsection{Databases}

As no public database is available for a tremor based person recognition purpose,  we have
collected our own dataset. For this purpose two smartphones, a Nexus 6 and a HTC One M9, both
running Android 5.0 Lollipop operating system, were used to construct two separate databases. The
application recorded the gyroscope and accelerometer data at a sample rate of 100 Hz. As the Nexus
database is larger (10 000 samples vs. 2150 samples), the discussed results will be based on this
database.

For the data collection, our goal was to collect the tremor data while the participants were
actively using the smartphone. During the data collection, most persons were using the developed
application for embedded camera. Some took pictures, others recorded videos, most people walked
while using the application. We have considered the possibility that the tremor fingerprints can be
different if the person is holding the smartphone with both hands or just one hand, so we have
asked the participants to vary their grip.

The training and the test data was gathered from 17 people. The data from a single person was
acquired in multiple sessions distributed among several weeks. For a person the recording session
took place at various moments of the day, to include examples where he is more relaxed or more
tired. The target group consisted of 9 males and 8 females in the [20-60] years range age.


\subsection{Training and Testing}

Data collected from all the subjects was pooled together. The train and test data was randomly drawn 
from this pool, ensuring no overlap between the train and test sets. We used this approach as it is
a good estimator for the generalized performance of the different classifiers.

Furthermore, to test in real-life condition, we ensure that in fact, for a single person, the
\emph{samples} used for training and respectively for testing were acquired in \emph{different
days}.

\paragraph{Accuracy measures.} In a biometric system, there are two types of errors \cite{jain2004}: 
(1) mistaking biometric measurements  from two different persons to be from the same person (called 
\emph{false match}) and (2) mistaking two biometric measurements from the same person to be from two 
different persons (called \emph{false nonmatch}). Also, the system performance at all the operating 
points (thresholds) can be depicted in the form of \emph{accuracy} which is the proportion of true
results (both true matches and true non-matches) amongst the total number of cases examined
{\cite{metz1978}}.


\subsection{Parameter Choice}

\paragraph{Classifier.} The family of decision trees contains multiple choice for classifiers. The
most popular are random forest and bagged ensemble of trees. To determine which choice yields the
best performance, we comparatively evaluated random forest, bagged ensemble and from other
categories, the K-nearest neighbour (KNN) based on Euclidean distance ($K=7$) and Support vector
Machine based on Sequential Minimal Optimization (SMO) learning algorithm. For the actual
implementation we relied on WEKA machine learning toolkit \cite{hall2009}. We have used 25\% of the
database for testing and we have assured that the data was recorded on different days. Results are
presented in Table \ref{tab:classifer}. As one may see, the Random Forest lead to best results.
Further experimentation rely on the Random Forest.

\begin{table}[t]
    \caption{Average accuracy, false match rate and false non-match rate for the classification algorithms tested with
    the full feature set with a one-second sliding window.}\label{tab:classifer} \centering
    \begin{tabular}{|m{1.5cm}|m{1.5cm}|m{1.4cm}|m{1.4cm}|}
        \hline
        \bf Classifier    & \bf Accuracy for the test data & \bf False Match Rate & \bf False Non-Match Rate \\ \hline
        Bagged Trees      & 0.70                           & 0.026                & 0.26                     \\ \hline
        \bf Random Forest & \bf 0.76                       & \bf 0.03             & \bf 0.24                 \\ \hline
        SMO               & 0.365                          & 0.079                & 0.63                     \\ \hline
        KNN               & 0.545                          & 0.036                & 0.49                     \\ \hline
    \end{tabular}
\end{table}

\begin{figure}[t]
    \vspace{-0.2cm}
    \centering
    \includegraphics[width=0.5 \textwidth]{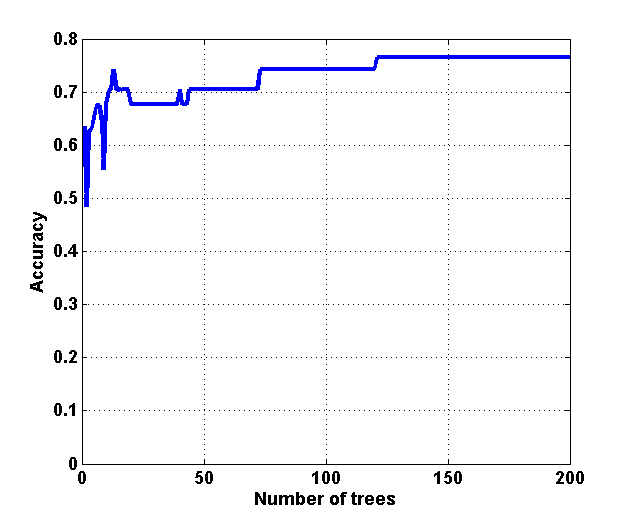}
    \caption{The classification accuracy increases with the tree number for the test data.}
    \label{fig:error}
    \vspace{-0.1cm}
\end{figure}

\paragraph{Number of trees.} We have experimented with one to 200 trees that were fully grown with
80\% of the attributes used in random selection. Ensembles tend to ``overtrain", meaning they
produce overly optimistic estimates of their predictive power, so we have tested how the number of
trees influences the classifier's accuracy on an independent test set (Figure \ref{fig:error}).
One may note that after 130 trees, the classification accuracy does not vary.

\paragraph{Window size.}
To determine the optimal sample length we experimented with different windows time sizes, from
0.5 to 5 seconds. The results showed in Table \ref{tab:win_size} indicate that the
one-second window has the best results, while the acquisition time is short enough not to be
perceived by the user.

Contrary to the standard disjoint windows we have also tried overlapping windows to various
percentages. Yet in all cases the accuracy decreased when compared to preferred choice.

\begin{table}[t]
    \caption{Average accuracy for the tested window sizes with the random classifier and 100 trees.}\label{tab:win_size} \centering
    \begin{tabular}{|m{2.1cm}|m{2cm}|m{2cm}|}
    	\hline
    	\bf Window size (sec) & \bf Average Accuracy & \bf Average accuracy\newline 50\% overlap \\ \hline
    	0.5                   & 0.477                & 0.506                                     \\ \hline
    	\bf 1                 & \bf 0.761            & 0.684                                     \\ \hline
    	1.5                   & 0.623                & 0.581                                     \\ \hline
    	2                     & 0.609                & 0.524                                     \\ \hline
    	3                     & 0.559                & 0.458                                     \\ \hline
    	5                     & 0.534                & 0.512                                     \\ \hline
    \end{tabular}
\end{table}


\subsection{Experiments}

\begin{figure}[t]
    \centering
    \includegraphics [width=0.45 \textwidth] {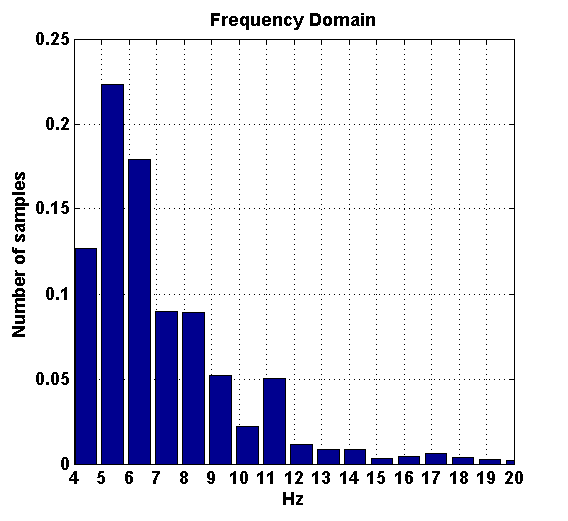}
    \caption{Frequency domain for the filtered hand tremor.}
    \label{fig:freq}

\end{figure}

\paragraph{Frequency Domain.} For the first experiment we aim to validate prior work conclusions
regarding relevant frequency domain for physiological tremor. Thus we computed and presented in
Figure \ref{fig:freq} the normalized power spectral density (i.e. frequency histogram). According
to our experiments 72\% of the tremor is contained by the to 4-7 Hz range, while 24\% corresponded
to 7-10 Hz and 45\% to 6-10 Hz.

The found results are in concordance with previous work \cite{andrade2013}, which concluded that the
bandwidth of the resting tremor is 4-7 Hz, as the bandwidth for the kinetic tremor is 7-15 Hz.

\paragraph{Feature ranking.} The tree ensemble classifier contains implicitly a feature ranking tool.
If one counts the number of splits based on a particular feature he will find the overall ranking.
For the proposed problem the most relevant features are presented in Table \ref{tab:feat_rank}.


\begin{table}[t]
    \caption{ Feature Ranking - the most relevant 5 features.}\label{tab:feat_rank} \centering
    \begin{tabular}{|m{0.7cm}|m{2.7cm}|m{2.7cm}|}
        \hline
        \bf Rank & \bf Feature                & \bf Dataset             \\ \hline
        1        & Irregularity J Periodogram & Accelerometer Magnitude \\ \hline
        2        & Irregularity K Periodogram & Accelerometer Z Axis    \\ \hline
        3        & Irregularity J Periodogram & Gyroscope Z Axis        \\ \hline
        4        & Irregularity K Periodogram & Accelerometer Magnitude \\ \hline
        5        & Irregularity K Periodogram & Gyroscope Y Axis        \\ \hline
    \end{tabular}
\end{table}

\begin{figure}[t]
    \centering
    {\includegraphics[width=0.5 \textwidth]{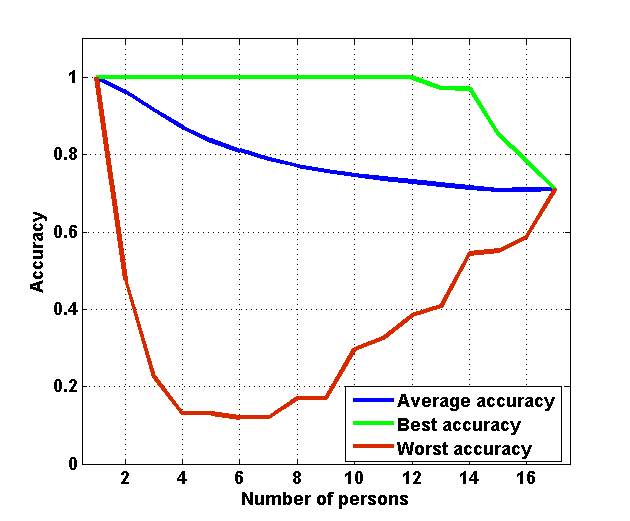}}
    \caption{Minimum, mean and maximum accuracy when the person number varies.}
    \label{fig:class}
\end{figure}

\paragraph{Performance across persons.}
Under the assumption that some individuals may have more similar features than others,  we have
trained and tested all the possible combinations of two to seventeen classes (i.e. persons). The
results are showed in Figure {\ref{fig:class}}. One should notice while the average performance is
always above 76\%, there are some classes that can be very similar (i.e. person cannot be
discriminated using the basic setup). Yet these results are obtained using the same set of features.
When we focused on the group of 4 that contained the ``similar'' persons, and we searched for a
different feature selection, the performance increased to over 60\%. Concluding, the discrimination
is still possible, but one needs to build a more powerful feature selection/machine learning
system.

\paragraph{Different smartphones.}
We have tried to determine if a person can be tracked using two smartphones.  In order to
test this situation and determine if this biometric can become a privacy issue, we have recorded
samples with both smartphones consecutively (i.e. database cross-validation). We have used the
classifier trained with the data gathered with the Nexus to classify the samples recorded with the
HTC and the average accuracy was 45\% significantly lower than 76\% achieved only on the Nexus
database. The performance evaluated only on the HTC, due to the fact that we were using a smaller database,
is even better (88\%).

Based on these findings, we conclude that the classifier (and the learned module) is currently  not
transferable from one smartphone to another. This may be explained based on the following findings:
(1) the most important features, as presented in Table \ref{tab:feat_rank}, are computed on the
accelerometer signal; (2) past work on device fingerprinting \cite{bojinov2014} shows that
accelerometers have a unique noise fingerprint. Thus the specificity off accelerometer noise is
incorporated with the learned model and currently it forces us to train the classifier
independently on each smartphone.

Tremor may not remain invariant over a long period of time, due to major injuries involving  hand
joints, certain medications, inebriety, muscle tiredness or weakness, normal aging, stress, anxiety
or fatigue. It may not be very distinctive, but it is sufficiently discriminatory for low-security
applications, like smartphone unlocking. This technique is very simple, relatively easy to use and
inexpensive, because it doesn't require additional sensors.

\paragraph{Verification.} We also tested the capabilities of the tremor system to work in the
verification mode. In this mode the system should validates a person's identity by comparing the
captured data with her own stored  template. Given the Nexus database of 17 persons with the
training scheme as discussed, we have run on each testing sample and we have achieved 98\% accuracy.

\section{Discussion}

\label{Sect:Discussion}

In this section we will review some of the found results and discuss some of the implications in
the background of general biometric measures.

\subsection{Does the tremor  qualify as a biometric measure?}

Jain et al. \cite{jain2004} notes that ``Any human physiological and/or behavioral characteristic
can be used as a biometric characteristic as long as it satisfies the following requirements'':
\emph{universality}, \emph{distinctiveness}, \emph{permanence}, \emph{collectability}.
Additionally, same the authors observe that \emph{performance},  \emph{acceptability}  and
\emph{circumvention} should be considered when discussing any biometric measure.

\paragraph{Universality.}
Previous works on the tremor \cite{mansur2007} \cite{andrade2013} concluded that it is present in
all persons. The main difference among different people is that certain ill persons exhibit an
augmented form of tremor (e.g. pathological tremor) characterized by lower dominant frequency and
increased amplitude. The finest form of tremor is encountered in individuals that actively practice
its reduction, such as microsurgeons; yet even here its amplitude does not go below the aimed 10
$\mu m$ \cite{Charles96} needed for most precise surgical operations.

\paragraph{Distinctiveness and performance.} This is the most obvious criteria when ranking various
biometric measures. The here computed performance is 76\% for different persons with a low for a
specific group of 4. In the worst case only after we used a dedicated feature selection we got a
60\% separation. Yet testing the capabilities of tremor as biometric measure is only at its initial
attempt and we are confident that dramatic increase of performance is achievable. To actually find
the measure limits, much larger databases (with more persons, different inertial sensors, many
realistic stresses) are needed.

However, it is also against intuition to believe that tremor may actively compete against biometrics
such as DNA or fingerprint in terms of accuracy. It is more likely that its distinctiveness may
outperform that of the gait and compete with written signature, thus falling into the ``low -
medium'' range.

\paragraph{Permanence.} Our tests showed that tremor characteristics are invariant over a period of
several weeks. Based on intuition one may argue that time invariance should span significantly
larger periods such as months to years. However aging and certain medical conditions (arm issues,
blood tensions, neurological diseases) will definitely affect it and limit its time invariance.

\paragraph{Collectability.} In this work we have used a smartphone that include inertial sensors to
acquire tremor data and rather simple and low-cost features to actively describe it. Furthermore,
to measure this biometric of one person he/she simply should hold in hand an device integrating
inertial sensors and recording means (analog-to-digital converter, processor, memory); the hardware
requirements are obviously easy to meet. In conclusion, we consider that tremor is highly collectible.

\paragraph{Acceptability.} As one simply has to hold something to have his/her tremor recorded, the
acquisition process is not invasive. Thus, our opinion is that tremor should have high acceptability
and we foresee this aspect will be one of the key feature that will lead to its development as a
biometric measure.

\paragraph{Circumvention.} Unfortunately, here we will count a weak point. The development of
piezoelectric transducer is at a level high enough, so that one will easily program such a device
to mimic another's tremor characteristics. We may only hope that in the development process of the
tremor as a biometric other features of the tremor, hard to be replicated, will be identified.

\subsection{Applications}

Typically, the biometric systems operate in either verification mode or identification mode. In
this paper we focused on the identification mode, reporting an average accuracy of 76\% and only
performed a side test for verification, which produced an accuracy of 98\%. However, we must point
out that the test for verification is only partially relevant as we tested the system only with
persons that were included in the training database. A completely relevant test for verification
should contain more persons in training and test on the tremor of completely new people.

Regarding practical applications, the most intuitive one is currently under development by us and
it is smartphone unlocking based on tremor. Given a trained system, after slightly more than one
second of holding, we are capable of reporting a result about holder's identity. This is faster,
less intrusive and less demanding than any PIN code introduction, swipe unlocking or face
recognition through front camera image processing.

Supplementary we anticipate that given initial person identification, afterwards the smartphone
can be customized to user preferences.


\subsection{Conclusion and future work}
Given the achieved performance of the random forests classifier the person recognition based on
tremor characteristics is indeed viable. Complementary, the hand tremor meets all the criteria for a
biometric measure. Thus the physiological tremor recognition can become a major security
improvement for smartphones with the added benefit that no additional hardware is needed.

To fully understand the tremor potential and limits as a biometric measure, significant additional
work is needed, spanning from larger database to trying other features for description and
different systems for classification.

Given the discrimination power amongst different persons, we consider that it is worthy to
investigate if the tremor may give indication about an unknown person's age, emotional state, grip
(and, consequently, posture); each of these suppositions, if validated, may end in a plethora of
practical applications. Furthermore, a tremor recognition system is easy to integrate not only into
a smartphone, but also into a smartwatch, a digital professional camera or it may be used in
automotive industry (as a initial security step).

\section*{\uppercase{Acknowledgements}}
This work was supported by the
Romanian National Agency for Scientific Research (MEN-UEFISCDI)
under the PNCDI2 no. 18/2014 (``Stabilization
Technology for Elimination of Absolute and DYnamic blur
due to Camera Acquisition Motion" -  SteadyCam) research
grant.

\bibliographystyle{icml2014}
\bibliography{HandTremor}

\end{document}